\documentclass{ws-ijmpe}

\newcommand{\ba}{\begin{eqnarray}}
\newcommand{\ea}{\end{eqnarray}}
\newcommand{\bmath}{\begin{subequations}}
\newcommand{\emath}{\end{subequations}}
\newcommand{\ban}{\begin{eqnarray*}}
\newcommand{\ean}{\end{eqnarray*}}
\newcommand{\tl}{\tilde{\ell}}
\begin{document}

\markboth{A. Leviatan}{Dirac-Symmetries: Doublet Structure and 
Supersymmetric Patterns}

%
\catchline{}{}{}{}{}
%

\title{
PSEUDOSPIN, SPIN, AND COULOMB DIRAC-SYMMETRIES:\\
DOUBLET STRUCTURE AND SUPERSYMMETRIC PATTERNS}

\author{A. LEVIATAN}

\address{Racah Institute of Physics, The Hebrew University,
Jerusalem 91904, Israel\\
ami@phys.huji.ac.il}

\maketitle

\begin{history}
\received{(received date)}
\revised{(revised date)}
\end{history}

\begin{abstract}
Relativistic symmetries of the Dirac Hamiltonian with a mixture
of spherically symmetric Lorentz scalar and vector potentials, 
are examined from the point of view of supersymmetric quantum mechanics. 
The cases considered include the Coulomb, pseudospin and spin limits 
relevant, respectively, to atoms, nuclei and hadrons.
\end{abstract}

\section{Introduction}

\begin{table}[pb]
\tbl{Doublet structure in the Coulomb, pseudospin, and spin limits of the 
Dirac equation. The notation $(n,\ell,j)$ refers, respectively, 
to the single-fermion radial, orbital, and angular momentum quantum 
numbers of the upper component of the Dirac wave function. 
The Dirac labels are 
$\kappa_1= -(j+1/2)$ and $\kappa_2=+(j^{\prime}+1/2)$.}
{\begin{tabular}{@{}cll@{}} \toprule
Limit & $\qquad\qquad\quad$ Doublet Structure & Dirac labels\\ \colrule
Coulomb & $(n,\ell,j=\ell +1/2) \quad$ 
$(n-1,\ell+1,j^{\prime}=\ell+1/2) \qquad $ & $\kappa_1 + \kappa_2 = 0$ \\
Pseudospin & $(n,\ell,j = \ell + 1/2)\quad$    
$(n-1,\ell + 2,j^{\prime}= \ell + 3/2) \qquad$ & $\kappa_1+\kappa_2 =1$ \\
Spin & $(n,\ell,j = \ell + 1/2)\quad $   
$(n,\ell,j^{\prime}= \ell - 1/2) \qquad$ & $\kappa_1+\kappa_2=-1$  \\
\botrule
\end{tabular}}
\end{table}
The Dirac equation serves as the basis for the 
relativistic description of atoms, nuclei and hadrons. 
In atoms the relevant potentials felt by the electron 
are Coulombic vector potentials. 
A Dirac Hamiltonian with a Coulomb potential 
exhibits a fine-structure spectrum with characteristic 
two-fold degeneracy. 
Relativistic mean fields in nuclei generated by 
meson exchanges~\cite{wal86}, 
and quark confinement in hadrons~\cite{gromes91} necessitate a mixture 
of Lorentz vector and scalar potentials. 
Recently symmetries of Dirac Hamiltonians with 
such Lorentz structure have been shown to be relevant for explaining 
the observed degeneracies of certain shell-model orbitals in nuclei 
(``pseudospin doublets'')~\cite{gino97}, 
and the absence of quark spin-orbit splitting 
(``spin doublets'')~\cite{page01}, 
as observed in heavy-light quark mesons. 
The degenerate doublets associated with the relativistic Coulomb, 
pseudospin, and spin symmetries are shown in Table 1.
In the current contribution we show~\cite{lev04} that 
the degeneracy patterns and relations between 
wave functions implied by such relativistic symmetries resemble 
the patterns found in supersymmetric quantum mechanics (SUSYQM). 
The feasibility of such a proposal 
gains support from the 
fact that the Dirac Hamiltonian with a vector Coulomb potential 
is known~\cite{suku85} to be supersymmetric. 

\section{Dirac Hamiltonian and Supersymmetric Quantum Mechanics}

The essential ingredients of 
SUSYQM~\cite{junker96} are the supersymmetric Hamiltonian 
${\cal H}~=~\left ({H_1\quad\atop 0}{0\atop H_2}\right )$ 
and charges 
$Q_{-} = \left ({ 0\atop L}{ 0\atop 0}\right )$, 
$Q_{+} = \left ({ 0\atop 0}{ L^{\dagger}\atop 0}\right )$ 
which generate the supersymmetry (SUSY) algebra
$[{\cal H},Q_{\pm}] = \{Q_{\pm},Q_{\pm}\}=0$, 
$\{Q_{-},Q_{+}\}= {\cal H}$. 
The partner Hamiltonians $H_1=L^{\dagger}L$ and $H_2=LL^{\dagger}$ 
satisfy an intertwining relation,  
\ba
LH_1 = H_2L ~,
\label{lh1h2}
\ea
where in one-dimension the transformation operator 
$L = \frac{d}{dx} + W(x)$ 
is a first-order Darboux transformation 
expressed in terms of a superpotential $W(x)$. The 
intertwining relation ensures that 
if $\Psi_1$ is an eigenstate of $H_1$, 
then also $\Psi_2=L\Psi_1$ is 
an eigenstate of $H_2$ with the same energy, 
unless $L\Psi_1$ vanishes or produces an unphysical state ({\it e.g.} 
non-normalizable). 
Consequently, as shown in Fig.~1(a), the SUSY 
partner Hamiltonians $H_1$ and $H_2$ are 
isospectral in the sense that their spectra 
consist of pair-wise degenerate levels 
with a possible non-degenerate single state in one sector (when 
the supersymmetry is exact). The wave functions of the degenerate levels 
are simply related in terms of $L$ and $L^{\dagger}$. 
Such characteristic features define 
a supersymmetric pattern. 
In what follows 
we focus the discussion on supersymmetric patterns 
obtained in selected Dirac Hamiltonians. 
\noindent
\begin{figure}[th]
\begin{minipage}{0.49\linewidth}
\centerline{\psfig{file=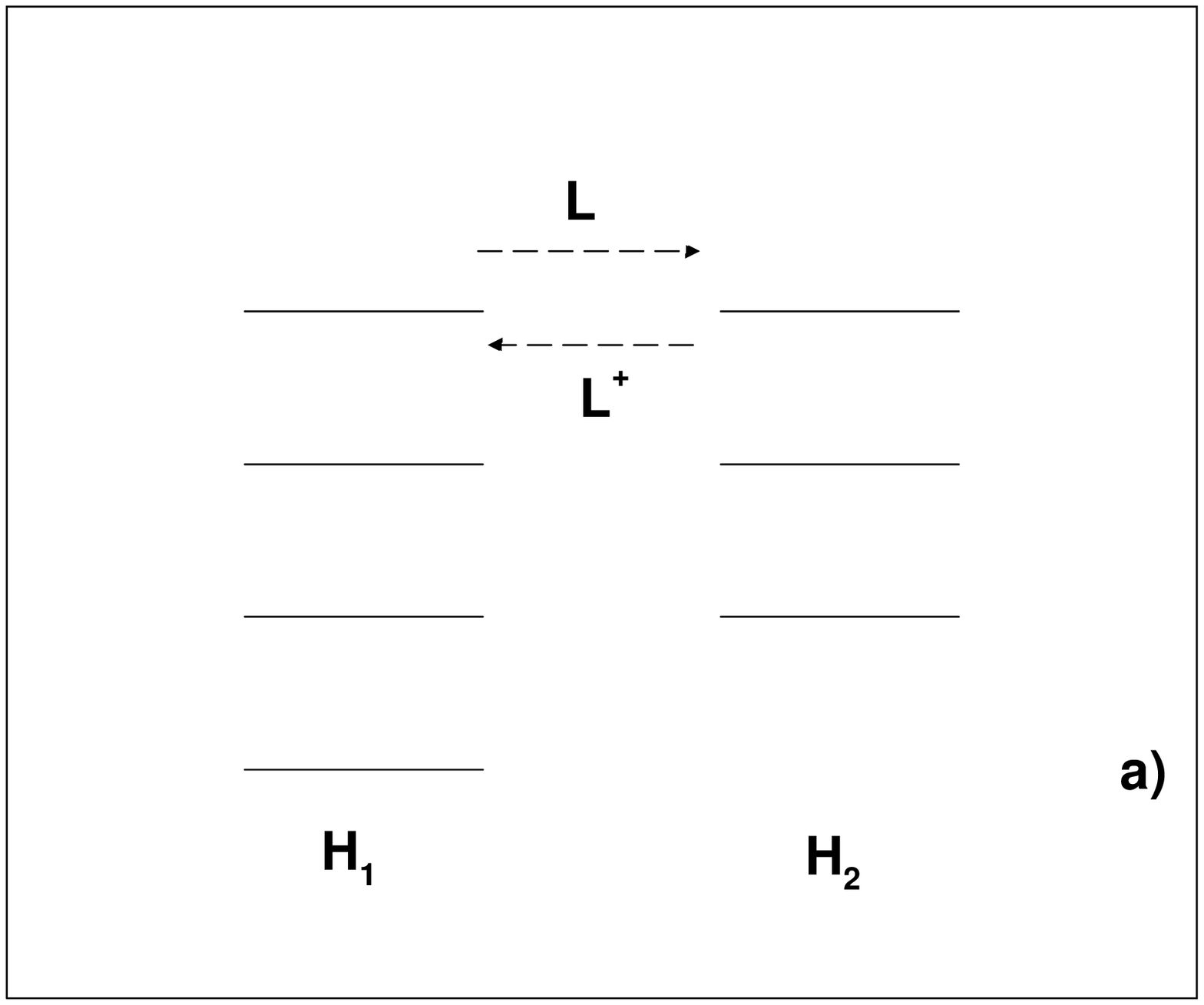,width=0.97\linewidth,angle=-90}}
\end{minipage}
\hspace{\fill}
\begin{minipage}{0.465\linewidth}
\vspace{-0.8cm}
\hspace{-0.2cm}
\centerline{\psfig{file=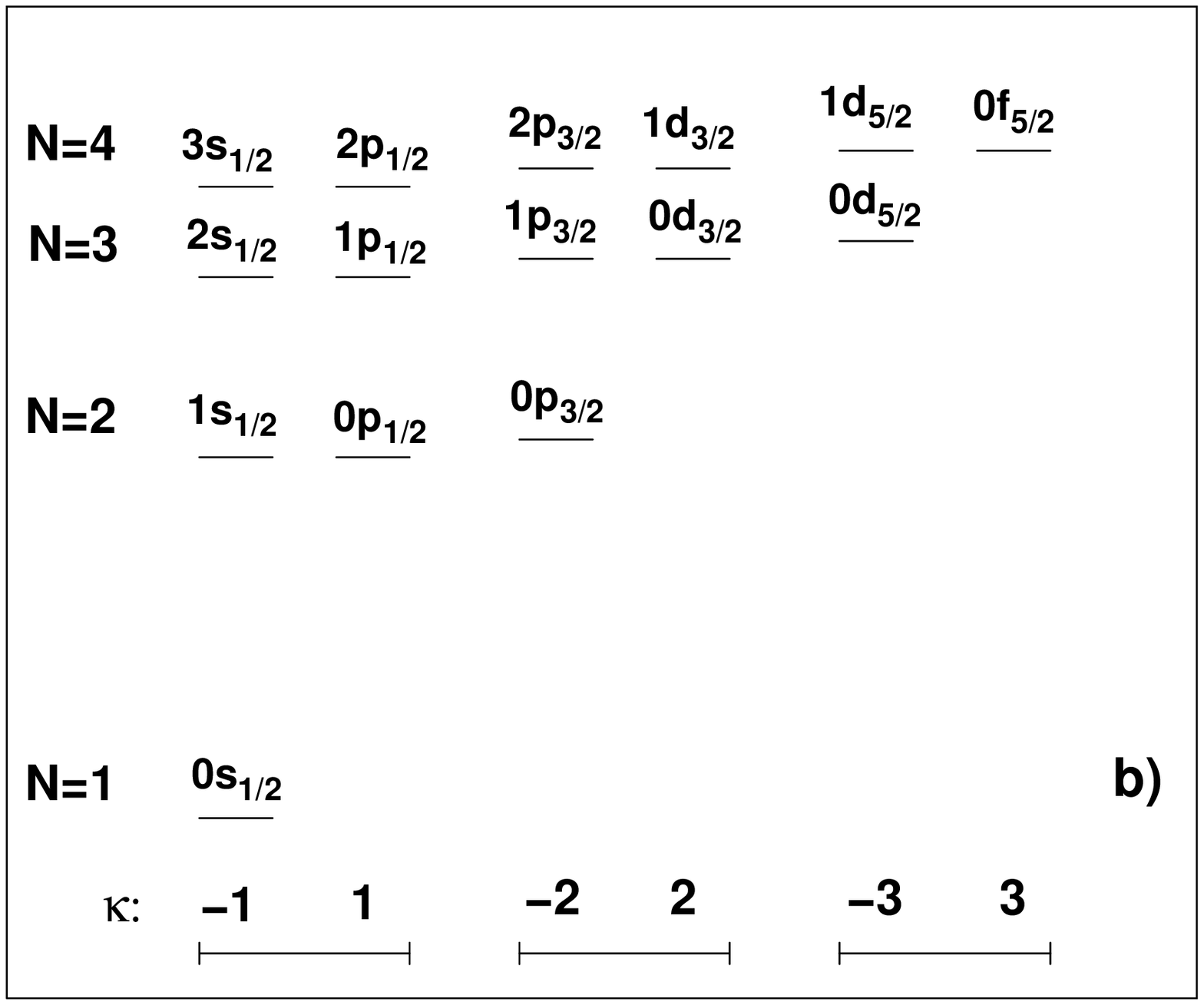,width=0.9\linewidth,angle=-90}}
\end{minipage}
\begin{minipage}{0.465\linewidth}
\vspace{-0.4cm}
\centerline{\psfig{file=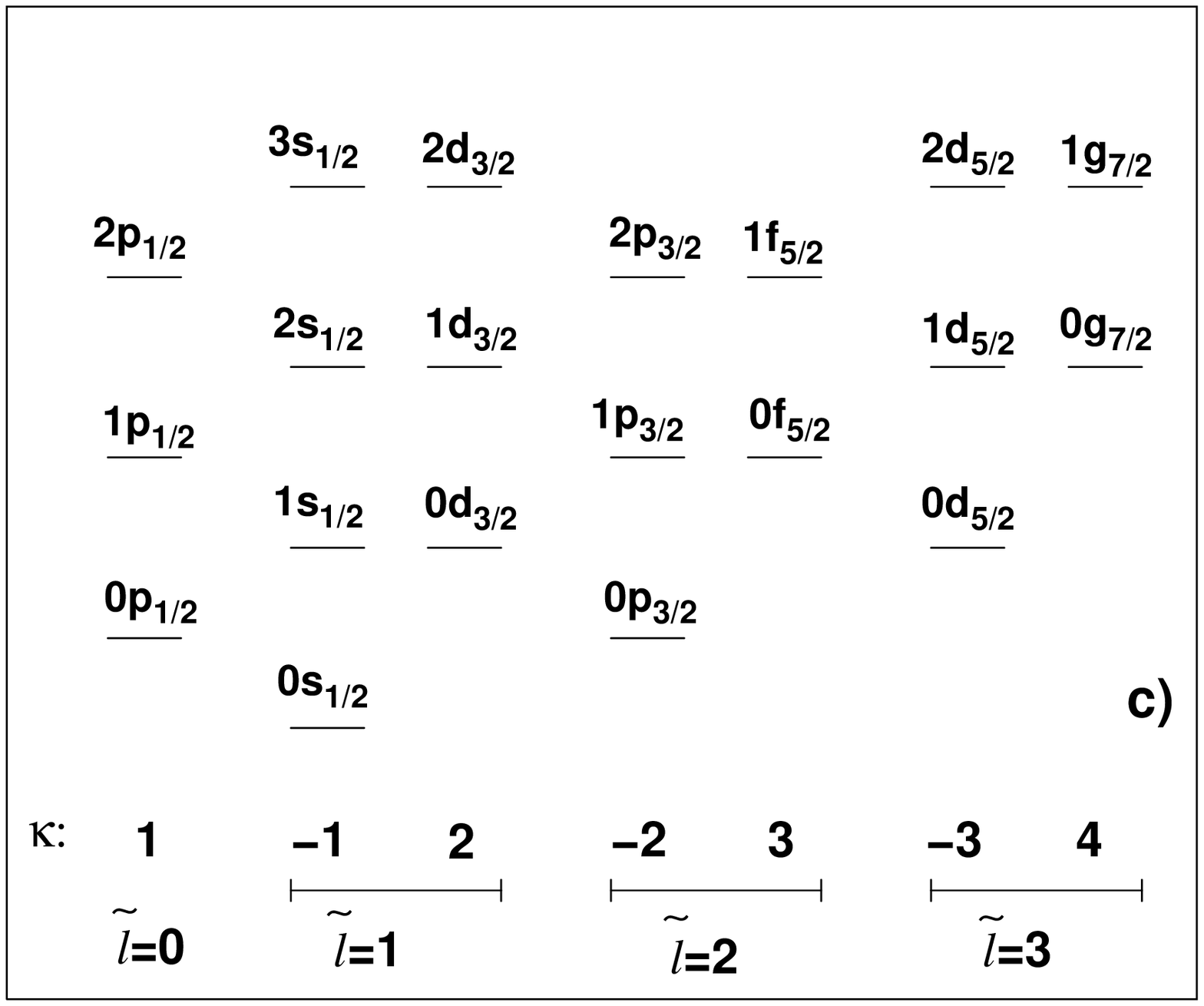,width=0.9\linewidth,angle=-90}}
\end{minipage}
\hspace{\fill}
\begin{minipage}{0.465\linewidth}
\vspace{-0.4cm}
\centerline{\psfig{file=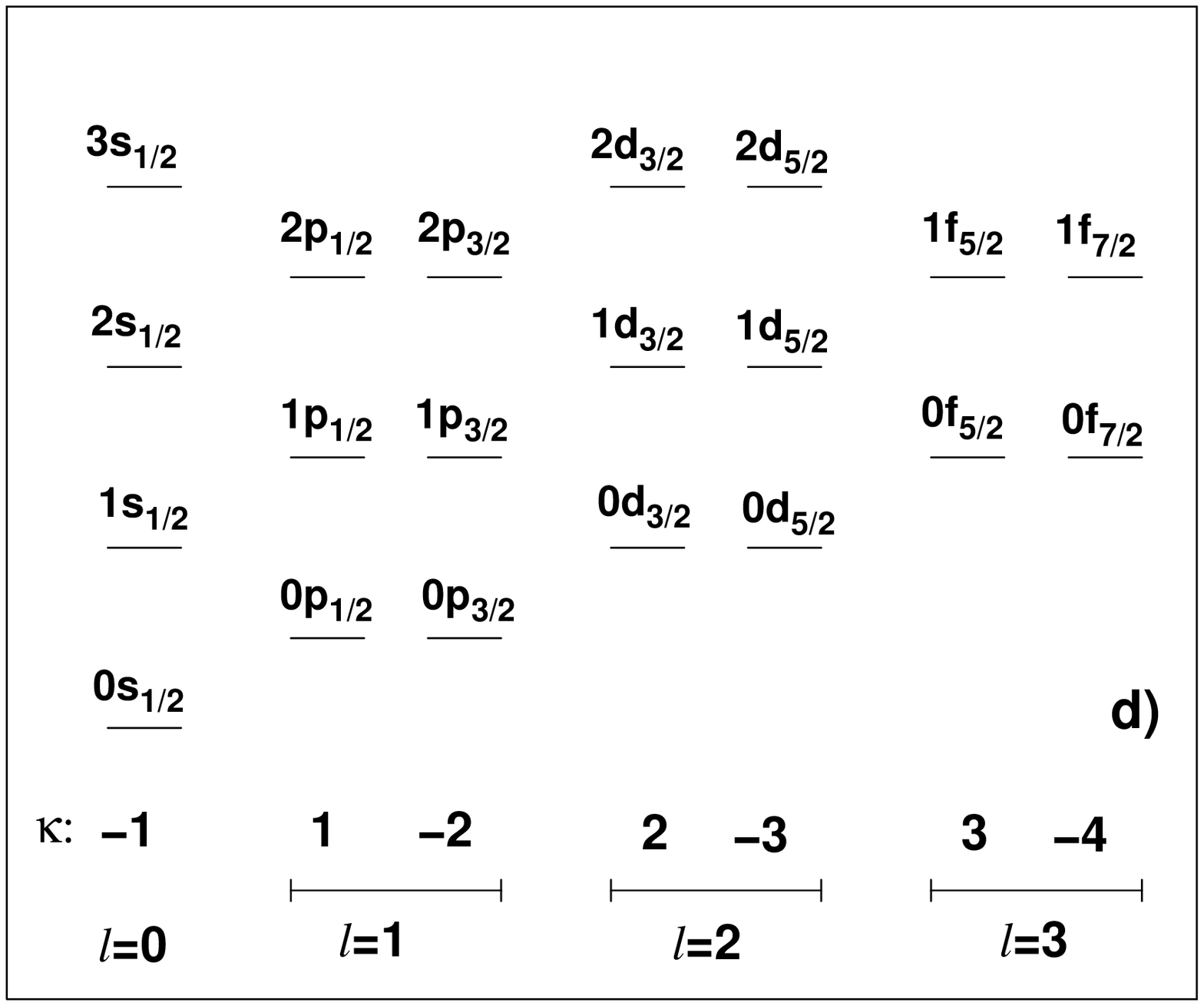,width=0.9\linewidth,angle=-90}}
\end{minipage}
\vspace*{8pt}
\caption{
Schematic supersymmetric patterns in (a) SUSYQM 
and in the (b) Coulomb, (c) pseudospin, (d) spin, limits 
of the Dirac Hamiltonian. 
In (a) $H_1$ and $H_2$ have identical spectra with an additional level 
for $H_1$ when SUSY is exact. 
Spectroscopic notation $n \ell j$ in (b)-(d) refers 
to quantum numbers of the upper component, and $\kappa$, $N$, $\tl$ 
are defined in the text. 
In (b) the radial nodes $n$ are related to $n_r$ by 
$n_r = n$ $(n_r = n +1)$ for $\kappa<0$ $(\kappa>0)$, and only the 
$E^{(+)}_{n_r,\kappa}$ branch is shown~\protect\cite{lev04}.}
\end{figure}

The Dirac Hamiltonian, $H$, for a fermion of mass~$M$ 
moving in external scalar, $V_S$, and vector,
$V_V$, potentials is given by 
$H = \hat{\bm{\alpha}}\bm{\cdot p}
+ \hat{\beta} (M  + V_S) + V_V$, 
where $\hat{\bm{\alpha}}$, 
$\hat{\beta}$ are the usual Dirac matrices 
and we have set the units $\hbar = c =1$. 
When the potentials are spherically symmetric: $V_S=V_S(r)$, $V_V=V_V(r)$, 
the operator 
$\hat{K} = -\hat{\beta}\,(
\bm{\sigma\cdot\ell} + 1)$, 
(with $\bm{\sigma}$ the Pauli matrices and 
$\bm{\ell} = -i\bm{r}\times \bm{\nabla}$), 
commutes with $H$ and 
its non-zero integer eigenvalues $\kappa = \pm (j+1/2)$ are used 
to label the Dirac wave functions
$\Psi_{\kappa,\,m} = 
r^{-1}
( G_{\kappa} [\,Y_{\ell}\,\chi\,]^{(j)}_{m},
iF_{\kappa}[\,Y_{\ell^{\prime}}\,\chi\,]^{(j)}_{m})$. 
Here $G_{\kappa}(r)$ and $F_{\kappa}(r)$ 
are the radial wave functions of the upper and lower components 
respectively, $Y_{\ell}$ and $\chi$ are the spherical harmonic and 
spin function which are coupled to angular momentum $j$ with 
projection $m$.
The labels $\kappa = -(j+1/2)<0$ and $\ell^{\prime}= \ell +1$ hold 
for aligned spin 
$j=\ell+1/2$ ($s_{1/2}, p_{3/2}$, etc.), while $\kappa = (j+1/2)>0$ 
and $\ell^{\prime}= \ell -1$ hold 
for unaligned spin $j=\ell-1/2$ 
($p_{1/2},d_{3/2},$ etc.). 
Denoting the pair of radial wave functions by
\ba
\Phi_{\kappa} = \left (
{G_{\kappa}\atop F_{\kappa}}
\right ) ~,
\label{radialwf}
\ea
the radial Dirac equations can be cast in Hamiltonian form, 
\ba
H_{\kappa}\,\Phi_{\kappa} &=& 
\left (
\begin{array}{ll}
M + \Delta   & \quad -\frac{d}{dr} + \frac{\kappa}{r} \\
\frac{d}{dr} + \frac{\kappa}{r} & \quad -(M + \Sigma) 
\end{array}
\right )
\left (
\begin{array}{c}
G_{\kappa}\\
F_{\kappa}
\end{array}
\right)
=E\,\Phi_{\kappa} ~,
\nonumber\\
\nonumber\\
\Delta(r) &=& V_S + V_V\;\ , \;\; \Sigma(r) = V_S - V_V.
\label{hkgen}
\ea
In analogy to Eq.~(\ref{lh1h2}) 
we now look for radial Dirac Hamiltonians  
$H_{\kappa_1}$ and $H_{\kappa_2}$ 
which satisfy an intertwining relation of the form 
\ba
LH_{\kappa_1} = H_{\kappa_2}L ~.
\label{lhk1hk2}
\ea
Following Ref.~[8] we consider a matricial Darboux 
transformation operator
\ba
L = A(r)\frac{d}{dr} + B(r) ~, 
\label{darboux}
\ea
where $A$ and $B$ are $2\times 2$ matrices with $r$-dependent entries 
$A_{ij}(r)$, $B_{ij}(r)$.
Relations~(\ref{lhk1hk2}) and~(\ref{darboux}) should be regarded as 
a system of equations for the unknown operator $L$ and the 
so-far unspecified potentials in 
$H_{\kappa}$ (\ref{hkgen}).
The transformation operator, when found, connects the two degenerate 
states 
\ba
L\,\Phi_{\kappa_1} = C\,\Phi_{\kappa_2}
\label{Lf1f2}
\ea
and imposes relations between their respective components.

In the usual application of SUSYQM, one starts from 
a solvable Hamiltonian $H_1$ and uses 
the intertwining relation to obtain 
a new solvable Hamiltonian $H_2$. In the present case we employ a different 
strategy, namely, insist that both partner Hamiltonians $H_{\kappa_1}$ and 
$H_{\kappa_2}$ be of the 
form prescribed in Eq.~(\ref{hkgen}) with the {\it same} potentials, 
and look for solutions of Eq.~(\ref{lhk1hk2}) such that the potentials 
are independent of $\kappa$. 
We find~\cite{lev04} three physically interesting 
solutions which require $\kappa_1+\kappa_2=0,1,-1$ and lead to the 
Coulomb, pseudospin, and spin limits respectively.

\section{The Coulomb limit $(\kappa_1+\kappa_2=0)$} 

The solution of Eq.~(\ref{lhk1hk2}) with $\kappa_1+\kappa_2=0$ 
fix the potentials to be of Coulomb type 
\ba
V_S &=& \frac{\alpha_{S}}{r} \;\; , \;\;
V_V = \frac{\alpha_{V}}{r} ~, 
\label{coulmbpot}
\ea 
(omitting constant shifts) with arbitrary strengths, 
$\alpha_S$, $\alpha_V$.
In terms of the quantities
$\eta_1 = (\alpha_S M + \alpha_V E)/\lambda \,$, 
$\eta_2 = (\alpha_S E + \alpha_V M)/\lambda\,$, 
$\lambda = \sqrt{M^2 - E^2}\,$, 
$\gamma = \sqrt{\kappa^2 + \alpha_{S}^2 - \alpha_{V}^2}\,$, 
the quantization condition reads: 
$\gamma + \eta_1 = -n_r$ 
$(n_r=0,1,2,\ldots\,)$, 
and leads to the 
eigenvalues
\ba
E^{(\pm)}_{n_r,\kappa}/M &=&
\frac{ -\alpha_{S}\alpha_{V} 
\pm (n_r+\gamma) \sqrt{\,(n_r+\gamma)^2 
-\alpha_{S}^2 + \alpha_{V}^2} }
{\alpha_{V}^2 + (n_r+\gamma)^2 } ~.
\ea
The $\kappa$-dependence enters through the factor $\gamma$. 
The spectrum consists of two branches denoted by superscript 
$(+)$ and $(-)$. 
The eigenfunctions are
\ba
\Phi_{n_r,\kappa} &=& 
\left (G_{\kappa}\atop F_{\kappa}\right ) = 
{\cal N}\, 
\left ({ -\sqrt{M+E}
[(\kappa+\eta_2)F_1 + n_r F_2]\atop 
\sqrt{M-E}\,
[(\kappa+\eta_2)F_1 - n_r F_2\,]}
\right )
\rho^{\gamma}e^{-\rho/2} ~, \nonumber\\
{\cal N} &=& \frac{\sqrt{\lambda}}{\Gamma(2\gamma+1)}
\left [\,\frac{\Gamma(2\gamma+n_r+1)}{2Mn_r!\,\eta_2(\kappa+\eta_2)}
\right ]^{1/2}
\label{coulombwf}
\ea
where $E = E^{(\pm)}_{n_r,\kappa}$ 
and $F_1=F(-n_r,2\gamma+1,\rho)$, $F_2 = F(-n_r+1,2\gamma+1,\rho)$ 
are confluent hypergeometric functions in the variable $\rho = 2\lambda r$. 
The states and energies in each branch are labeled by $(n_r,\kappa)$. 
It is also possible to express the results in terms of the principal quantum 
number $N$ defined as $N= n_r + \vert \kappa\vert$, $(N=1,2,\ldots\,)$. 
For $n_r\geq 1$ 
the eigenvalues in each 
branch are two-fold degenerate with respect to the sign of 
$\kappa$, {\it i.e.} 
$E^{(+)}_{n_r,\kappa}=E^{(+)}_{n_r,-\kappa}$ and 
$E^{(-)}_{n_r,\kappa}=E^{(-)}_{n_r,-\kappa}$. 
For $n_r=0$ there is only {\it one} acceptable state for each $\kappa$, 
which has $\kappa<0$ for the $(+)$ branch and 
$\kappa >0$ for the $(-)$ branch. Equivalently, 
for a fixed principal quantum number $N$, 
the allowed values of $\kappa$ 
are $\kappa = \pm 1, \pm 2,\ldots, \pm (N-1),-N$ for the 
$(+)$ branch and $\kappa = \pm 1, \pm 2,\ldots, \pm (N-1),+N$ for the 
$(-)$ branch of the spectrum.

Focusing on the set of states with $\kappa_1=-\kappa_2\equiv \kappa$, 
the levels are separated according to the value of 
$\vert\kappa\vert = j+1/2$. 
For fixed $\kappa$, $E^{(+)}_{n_r,\kappa}$ is an increasing function 
of $n_r$ and, as shown in Fig.~1(b), 
for each value of $j$ we have a characteristic supersymmetric pattern.
There are two 
towers of energy levels, 
one for $-\vert\kappa\vert$ (with $n_r=0,1,2,\dots$) and one for 
$+\vert\kappa\vert$ (with $n_r=1,2,\ldots$). 
The two towers are identical, except that the 
$E^{(+)}_{n_r=0,-\vert\kappa\vert}$ 
level at the bottom of the $-\vert \kappa\vert$ 
tower is non-degenerate. 
Similar patterns of pair-wise degenerate levels with $\pm\kappa$ 
appear also in the $(-)$ branch of the spectrum. 
However, since for fixed $\kappa$, 
$E^{(-)}_{n_r,\kappa}$ is a decreasing function of $n_r$, 
the non-degenerate $E^{(-)}_{n_r=0,\vert\kappa\vert}$ level is now 
at the top of the $+\vert \kappa\vert$ tower, 
resulting in an inverted supersymmetric pattern. 
The transformation operator is given by 
\ba
L = a \left (
\begin{array}{cc}
\frac{d}{dr} +\frac{\epsilon_{+}}{r} + \frac{M\alpha_{+}}{\kappa_1}  &  
-\frac{\alpha_S}{\kappa_1}\frac{d}{dr} + \frac{\alpha_V}{r} \\
\frac{\alpha_S}{\kappa_1}\frac{d}{dr} - \frac{\alpha_V}{r} &
\frac{d}{dr} -\frac{\epsilon_{-}}{r} - \frac{M\alpha_{-}}{\kappa_1}
\end{array}
\right ) ~, 
\label{Lcoulomb}
\ea 
where $\epsilon_{\pm} = \kappa_1 
+ \alpha_S\alpha_{\pm}/\kappa_1$ 
and $\alpha_{\pm}= (\alpha_S \pm \alpha_V)$. 
Relation (\ref{Lf1f2}) is satisfied with 
$C =\frac{a\lambda}{\kappa_1}\sqrt{n_r(\gamma -\eta_1)}$ 
and $\kappa_1+\kappa_2=0$. 
The operator $L$ connects degenerate 
states with $(n_r\geq 1,\pm\kappa)$, 
and annihilates the non-degenerate states with $n_r=0$.
The condition $\kappa_1+\kappa_2=0$ determines the angular momentum 
couplings in the full Dirac wave functions of the doublet 
\ba
\Psi_{\kappa_1<0,\,m} = 
\frac{1}{r} \left (
G_{\kappa_1} [\,Y_{\ell}\,\chi\,]^{(j)}_{m} 
\atop
iF_{\kappa_1}[\,Y_{\ell+1}\,\chi\,]^{(j)}_{m}
\right) \qquad
\Psi_{\kappa_2>0,\,m} =
\frac{1}{r} \left (
G_{\kappa_2} [\,Y_{\ell+1}\,\chi\,]^{(j^{\prime})}_{m}
\atop
iF_{\kappa_2}[\,Y_{\ell}\,\chi\,]^{(j^{\prime})}_{m}
\right) \;
\label{wfcoul}
\ea
where $\kappa_1 = -(\ell+1)<0,\; j=\ell +1/2$ and 
$\kappa_2= +(\ell+1)>0,\; j^{\prime}=\ell+1/2$.

The explicit solvability and observed degeneracies of the relativistic 
Coulomb problem are related to the existence of an additional conserved 
Hermitian operator~\cite{lev04} 
\ba
B= -i\hat{K}\gamma_5\,\left (H - \hat{\beta}M\right ) + 
\frac{\bm{\sigma \cdot r}}{r} 
\left 
(\,\alpha_{V} M + \alpha_{S} H\,\right ) ~, \;\;\; 
\label{JL}
\ea
which commutes with the full Dirac scalar and vector Coulomb Hamiltonian, 
$H$, but anticommutes with $\hat{K}$.
This operator is a generalization of the 
Johnson-Lippmann operator~\cite{john50} applicable for $\alpha_S=0$. 

\section{The pseudospin limit $(\kappa_1+\kappa_2 =1)$}

The solution of Eq.~(\ref{lhk1hk2}) with $\kappa_1+\kappa_2=1$ requires 
that the sum of scalar and vector potentials is a constant 
\ba
\Delta(r) = V_{S}(r) + V_{V}(r) = \Delta_0 ~.
\label{pspot}
\ea 
In this case the transformation operator is given by 
\ba
L = b \left (
\begin{array}{cc}
0 \quad & \frac{d}{dr} - \frac{\kappa_2}{r} \\
-\frac{d}{dr} - \frac{\kappa_1}{r} \quad & 
(2M + \Sigma + \Delta_0)
\end{array}
\right ) ~. 
\label{Ldel0}
\ea
Relation (\ref{Lf1f2}) is obeyed with
$\kappa_1+\kappa_2=1$, 
$C = b(M + \Delta_0 - E)$, $E=E_{\kappa_1}=E_{\kappa_2}$ 
and, consequently, the radial components satisfy 
\ba
\frac{dG_{\kappa_1}}{dr} + \frac{\kappa_1}{r}G_{\kappa_1} 
&=& \frac{dG_{\kappa_2}}{dr} + \frac{\kappa_2}{r}G_{\kappa_2} ~,
\nonumber\\
F_{\kappa_1} &=& F_{\kappa_2} ~.
\label{GFps}
\ea
The condition $\kappa_1+\kappa_2=1$ determines the form of the full 
Dirac wave functions of the doublet states 
\ba
\Psi_{\kappa_1<0,\,m} = 
\frac{1}{r} \left (
G_{\kappa_1} [\,Y_{\ell}\,\chi\,]^{(j)}_{m} 
\atop
iF_{\kappa_1}[\,Y_{\ell+1}\,\chi\,]^{(j)}_{m}
\right) \qquad
\Psi_{\kappa_2>0,\,m} =
\frac{1}{r} \left (
G_{\kappa_2} [\,Y_{\ell+2}\,\chi\,]^{(j^{\prime})}_{m}
\atop
iF_{\kappa_2}[\,Y_{\ell+1}\,\chi\,]^{(j^{\prime})}_{m}
\right) \;
\label{wfps}
\ea
where $\kappa_1 = -(\ell+1) <0,\; j=\ell+1/2$ and 
$\kappa_2= \ell+2 >0,\; j^{\prime}=\ell+3/2$.
From Eqs.~(\ref{GFps})-(\ref{wfps}) we see that the lower components 
of the two states in the doublet have the same spatial wave function. 
In particular, they have the same orbital angular momentum $\tl=\ell+1$,
and the same number of nodes, $n$. Eq.~(\ref{wfps}) 
and a theorem~\cite{levgino01} concerning the 
nodal structure of Dirac bound states ensure that the corresponding 
upper components have quantum numbers $(n,\ell,j = \ell + 1/2)$ for 
$\kappa_1<0$ and $(n-1,\ell + 2,j = \ell + 3/2)$ for $\kappa_2>0$. 
These are precisely the identifying features of 
pseudospin doublets~\cite{arima69} in nuclei. 
The latter refer to the empirical observation 
of quasi-degenerate pairs of normal-parity
shell-model orbitals with such non-relativistic quantum numbers. 
The doublet structure is expressed in terms of the ``pseudo'' orbital 
angular momentum, $\tilde{\ell}$ = $\ell+1$, and ``pseudo'' 
spin, $\tilde{s} = 1/2$, which are coupled to 
$j = \tilde{\ell}\pm \tilde s$. 
Such doublets play a central role in explaining features of 
nuclei~\cite{bohr82}, including superdeformation and identical bands. 
For potentials with asymptotic behavior as encountered in nuclei, 
the Dirac eigenstates for which both the 
upper ($G_{\kappa}$) and lower ($F_{\kappa}$) components have no nodes, 
can occur~\cite{levgino01} only for $\kappa <0$, and hence do not have 
a degenerate partner eigenstate (with $\kappa >0$). 
These nodeless Dirac states correspond to the shell-model states 
with $(n~=~0,\ell,j=\ell+1/2)$. 
For heavy nuclei such states with 
large $j$, {\it i.e.}, $0g_{9/2},\;0h_{11/2},\;0i_{13/2}$, 
are the ``intruder'' abnormal-parity states which, indeed, empirically 
are found not to be part of a doublet~\cite{bohr82}. 
Altogether, as shown in Fig.~1(c), the ensemble of Dirac 
states with $\kappa_1+\kappa_2=1$ 
exhibits a supersymmetric pattern of twin towers with pair-wise degenerate 
pseudospin doublets sharing a common $\tl$, and an additional 
non-degenerate nodeless state at the bottom of the $\kappa_1<0$ tower. 
An exception to this rule 
is the tower with $\kappa_2=1$ ($p_{1/2}$ states with $\tl=0$), 
which is on its own, because states with $\kappa_1=0$ do not exist. 

For potentials satisfying the condition of Eq.~(\ref{pspot}), 
the Dirac Hamiltonian is invariant under an SU(2) 
algebra, whose generators are~\cite{bell75,ginolev98} 
\ba
{\hat{\tilde {S}}}_{\mu} =
\left (
\begin{array}{cc}
\hat {\tilde s}_{\mu} &  0 \\
0 & {\hat s}_{\mu}
\end{array}
\right ) ~.
\label{Sgen}
\ea
Here 
${\hat s}_{\mu} = \sigma_{\mu}/2$ are the usual spin 
operators, $\hat {\tilde s}_{\mu}= U_p {\hat s}_{\mu} U_p$ 
and $U_p = \frac{\bm{\sigma\cdot p}}{p}$. 
This relativistic symmetry has been used~\cite{gino97} 
to explain the occurrence 
of pseudospin doublets in nuclei. 
Eqs.~(\ref{GFps})-(\ref{wfps}) are derived by 
exploiting the fact that in the symmetry limit the Dirac eigenfunctions 
belong to the spinor representation of the pseudospin SU(2) algebra.
The relations in Eq.~(\ref{GFps}) between the radial 
components of the doublet wave functions, 
have been tested in numerous 
realistic mean field calculations in a variety of nuclei, 
and were found to be obeyed to a good approximation, especially for 
doublets near the Fermi surface~\cite{ginomad98,ginolev01}. 
By the above discussion these results confirm the relevance of 
supersymmetry to single-particle states in nuclei.

\section{The spin limit $(\kappa_1+\kappa_2 =-1)$}

The solution of Eq.~(\ref{lhk1hk2}) with $\kappa_1+\kappa_2=-1$ 
requires that the difference of the scalar and vector potentials is a 
constant 
\ba
\Sigma(r) = V_{S}(r) - V_{V}(r) = \Sigma_0 ~.
\label{sig0}
\ea
The transformation operator is given by
\ba
L = -b \left (
\begin{array}{cc}
(2M + \Sigma_0 + \Delta ) \quad & 
-\frac{d}{dr} + \frac{\kappa_1}{r} \\
\frac{d}{dr} + \frac{\kappa_2}{r} \quad & 0  
\end{array}
\right ) ~.
\label{Lsig0}
\ea 
It connects the two doublet states 
as in Eq.~(\ref{Lf1f2}) with $\kappa_1+\kappa_2=-1$, 
$C = -b(E + M + \Sigma_0)$ and $E=E_{\kappa_1}=E_{\kappa_2}$. The 
corresponding radial components then satisfy 
\ba
G_{\kappa_1} &=& G_{\kappa_2}
\nonumber\\
\frac{dF_{\kappa_1}}{dr} - \frac{\kappa_1}{r}F_{\kappa_1} &=& 
\frac{dF_{\kappa_2}}{dr} - \frac{\kappa_2}{r}F_{\kappa_2}~. 
\label{GFss}
\ea
The condition $\kappa_1+\kappa_2=-1$ determines the form of the full 
Dirac wave functions of the two states in the doublet 
\ba
\Psi_{\kappa_1<0,\,m} = 
\frac{1}{r} \left (
G_{\kappa_1} [\,Y_{\ell}\,\chi\,]^{(j)}_{m} 
\atop
iF_{\kappa_1} [\,Y_{\ell+1}\,\chi\,]^{(j)}_{m}
\right) \qquad 
\Psi_{\kappa_2>0,\,m} =
\left (
G_{\kappa_2} [\,Y_{\ell}\,\chi\,]^{(j^{\prime})}_{m}
\atop
iF_{\kappa_2} [\,Y_{\ell-1}\,\chi\,]^{(j^{\prime})}_{m}
\right) \;
\label{wfss}
\ea
where $\kappa_1 = -(\ell+1)$, $j=\ell+1/2$ and
$\kappa_2 = +\ell$, $j^{\prime}=\ell-1/2$. 
Using Eq.~(\ref{GFss}) we see that the upper components 
in Eq.~(\ref{wfss}) share the same spatial wave function, and have 
quantum numbers $(n,\ell,j = \ell + 1/2)$ for $\kappa_1<0$ and 
$(n,\ell,j = \ell - 1/2)$ for $\kappa_2>0$. 
As shown in Fig.~1(d), the spectrum consists of towers of states with 
$\kappa_1+\kappa_2=-1$, forming pair-wise degenerate spin doublets. 
In this case, none of the towers have a single non-degenerate state 
and hence, the spectrum corresponds to that of a 
broken SUSY~\cite{junker96}. 
The tower with $\kappa_1=-1$ ($s_{1/2}$ states) is on its own, since 
states with $\kappa_2=0$ do not exist. 

For potentials satisfying condition (\ref{sig0}) the Dirac Hamiltonian 
is invariant under another SU(2) 
algebra, whose generators are obtained from Eq.~(\ref{Sgen}) by 
interchanging ${\hat s}_{\mu}$ and $\hat {\tilde s}_{\mu}$~\cite{bell75}
\ba
\hat{S}_{\mu} =
\left (
\begin{array}{cc}
{\hat s}_{\mu} & 0\\
0 & \hat {\tilde s}_{\mu}
\end{array}
\right ) ~.
\label{Spgen}
\ea
Eqs.~(\ref{GFss})-(\ref{wfss}) follow from the fact that 
the Dirac eigenfunctions in the spin limit are spinors with respect 
to this SU(2) algebra. The spin doublets resulting from 
this relativistic symmetry were argued to be relevant for 
degeneracies observed in heavy-light quark mesons~\cite{page01}.

\section{Summary}

We have examined three symmetry limits of a Dirac Hamiltonian with 
spherically-symmetric scalar and vector potentials, from a 
supersymmetric quantum mechanics perspective. 
In the Coulomb limit the potentials are $1/r$ but their strengths are 
otherwise arbitrary. In the pseudospin or spin limits there are no 
restrictions on the $r$-dependence of the potentials but there is a 
constraint on their sum or difference. These relativistic symmetries 
lead to degenerate doublets with quantum numbers shown in Table 1, 
and impose relations between the respective doublet wave functions. 
The latter relations are precisely the conditions needed for the 
fulfillment of an intertwining relation, Eq.~(\ref{lhk1hk2}), 
which is the underlying mechanism of SUSYQM.
The resulting supersymmetric patterns exhibit sectors of pair-wise 
degenerate doublets, with a possible nondegenerate single state in one 
sector. It is gratifying to note that the indicated supersymmetric patterns 
are manifested empirically, to a good approximation, in physical dynamical 
systems. While previous studies have focused on properties of 
individual doublets in nuclei and hadrons, 
it is the grouping of several doublets (and intruder levels in nuclei) 
into larger multiplets, as discussed in the present contribution, 
which highlights the fingerprints of supersymmetry 
present in these dynamical systems.

\section*{Acknowledgements}

It is a pleasure and privilege to dedicate this article to 
Bruce Barrett on the occasion of his 60th birthday.
This work was supported by the Israel Science Foundation.

\end{document}